\documentclass[aps,twocolumn,superscriptaddress,footinbib,floatfix,showpacs]{revtex4-1} 
\usepackage{amsfonts}
\usepackage{amssymb}
\usepackage[fleqn]{amsmath}
\usepackage[mathscr]{eucal}
\usepackage{graphicx} 
\usepackage{subfigure}

\begin{document} 

\title{Competitively Coupled Maps and Spatial Pattern Formation}

\author{Timothy Killingback}
\affiliation{Department of Mathematics, University of Massachusetts, Boston, Massachusetts 02125.}    
\author{Gregory Loftus} 
\affiliation{Department of Physics, University of Massachusetts, Boston, Massachusetts 02125.}    
\author{Bala Sundaram}
\affiliation{Department of Physics, University of Massachusetts, Boston, Massachusetts 02125.}

\begin{abstract}
Spatial pattern formation is a key feature of many natural systems in physics, chemistry and biology. The essential theoretical issue in understanding
pattern formation is to explain how a spatially homogeneous initial state can undergo spontaneous symmetry breaking
leading to a stable spatial pattern. This problem is most commonly studied using partial differential equations to model
a reaction-diffusion system of the type introduced by Turing. We report here on a much simpler and more robust model of spatial pattern
formation, which is formulated as a novel type of coupled map lattice. In our model, the local site dynamics are coupled
through a competitive, rather than diffusive, interaction. Depending only on the strength of the interaction, this competitive
coupling results in spontaneous symmetry breaking of a homogeneous initial configuration and the formation of stable spatial patterns. This mechanism
is very robust and produces stable pattern formation for a wide variety of spatial geometries, even when the local site dynamics
is trivial.
\end{abstract}
\maketitle

\section{INTRODUCTION}

Achieving a satisfactory understanding of spatial pattern formation is an enduring problem in the physical and biological sciences~\cite{Giereretal}.
At the fundamental level the essential issue is to explain how a spatially inhomogeneous stationary pattern can emerge dynamically from a
spatially homogeneous initial state. The most widely studied mechanism for producing such spatial patterns is that proposed by Turing~\cite{Turing}.
Turing's model is formulated as a system of coupled reaction-diffusion equations that describe the dynamical interaction of two chemical species
or morphogens. Pattern formation in Turing's system results from the counterintuitive fact that, under suitable conditions, the diffusion of the
morphogens can drive symmetry breaking in the initial homogeneous configuration. This is possible in Turing's system because diffusion is acting
in concert with other processes: namely, one of the morphogens (which undergoes an autocatalytic reaction) activates the other morphogen, while 
this morphogen in turn inhibits the former. Whether such a system can produce stationary spatial patterns (as opposed to, for example, spiral
waves as in the Belosov-Zhabotinsky reaction~\cite{BZ}) depends delicately on the diffusion rates of the morphogens, and on the rate at which the inhibitor
responds to changes in the activator concentration~\cite{Meinh}. This subtle dependence of the mechanism on parameter values makes Turing structures
challenging to produce experimentally~\cite{Castit}, which suggests that the mechanism may have limited applicability to pattern formation in
many real-world situations.

A variety of spatio-temporal patterns have also been shown to arise in the context of coupled-map lattices. Originally introduced by
Kaneko, these well-studied systems~\cite{Kaneko} consider a lattice geometry where the temporal site dynamics of a single variable is described by a discrete
time mapping, while a functional dependence on the neighboring values of the site variable determines the spatial dynamics. This coupling between
sites is often diffusive in character~\cite{Kaneko}, though a number of variants have been studied (see for example Ref.~\cite{Keel}). 
A common feature of these models is that the uncoupled site 
dynamics needs to be sufficiently complex (often chaotic in terms of the local map dynamics) in order to generate non-trivial spatial patterns on 
introducing spatial interactions. As is clear from the case of diffusive coupling, the spatial coupling serves to coarse-grain the large 
variations at the sites leading to patterns. A natural question that proceeds from this is whether or not there are mechanisms to generate 
non-trivial spatial patterns when the on-site dynamics is fully trivial (i.e. the local dynamics has a stable equilibrium point)? 

The purpose of this paper is to study a model of spatial pattern formation 
which is both simpler and more robust than
Turing-type models and readily produces stationary patterns from a homogeneous initial configuration.
Our model is formulated as a variant of a coupled map lattice~\cite{Kaneko}, in which there is a single dynamical variable (i.e. a single species or morphogen level) 
at each spatial site and
the site dynamics are coupled through a competitive (that is, inhibitory) interaction rather than through a conventional diffusive
coupling. The competitive coupling in our model allows the production of complex spatial patterns even when the dynamics at
each site is trivial, in the sense that the local dynamics exhibits a stable fixed point. This is in stark contrast to the dynamical
behavior in conventional diffusively coupled map lattices where
trivial site dynamics can only result in a spatially homogeneous state~\cite{WallKap,LJans,Jack90,Roh96}. This latter result reflects the fact that for
a single species model diffusion can never have a destabilizing effect. In diffusively coupled map lattices, pattern formation can occur but this
is only possible when the local site dynamics is complex. Indeed, non-trivial pattern formation typically requires that the local
dynamics is chaotic.~\cite{Kaneko}

In order to appreciate the importance of competitive coupling, as opposed to diffusive coupling, for spatial pattern formation in map lattices, it is necessary to briefly review the stability properties of diffusively coupled map lattices. An important issue in formulating a diffusively coupled map lattice is the form taken for the diffusive coupling term. The initial work on coupled map lattices~\cite{Kaneko,WallKap} considered maps of the form 
\begin{equation}
x_i(t+1)=f(x_i(t))+d[x_{i+1}(t)+x_{i-1}(t)-2x_i(t)],
\label{cml1}
\end{equation}
in one-dimension with nearest neighbor interactions, and the obvious generalizations to other spatial dimensions and interaction neighborhoods. The map $f(x)$ defines the local dynamics, and $d$ represents a ``diffusion'' constant. The case in which $f(x)$ is the logistic map $f(x)=rx(1-x)$ has often been considered in the literature~\cite{Kaneko,WallKap,Kap85}. The question of whether non-trivial spatial patterns can arise when the local dynamics is trivial (i.e., when the local dynamics $f(x)$ has a stable fixed point) depends on the stability of the coupled map (\ref{cml1}). For $f(x)$ of the form of the logistic map the stability theory of (\ref{cml1}) has been derived in Ref.~\cite{WallKap}, and also in Ref.\cite{Jack90}. It has been shown that (\ref{cml1}) always has a stable homogeneous fixed point if the map $f(x)$ has a stable fixed point. Thus, if the local dynamics is trivial then it is impossible to obtain non-trivial spatial pattern formation in the coupled map lattice. For this reason attempts to find non-trivial patterns in coupled maps of this form have used local maps with non-trivial dynamics. For example, in Ref.~\cite{Kap85} spatial pattern formation was found in the two-dimensional version of (\ref{cml1}) with $f(x)=rx(1-x)$, and $r=3.25$ and $r=3.30$, which correspond in both cases to the local dynamics being a stable 2-cycle. 

The coupled map lattice (\ref{cml1}) has, however, a fundamental problem, which was first identified in ~\cite{Yam83} (see also ~\cite{Jack90}): the coupled map (\ref{cml1}) is mathematically not well-defined --- namely, the iterates of (\ref{cml1}) can become unbounded for perfectly reasonable values of $d$ and initial values $x_i (0)$. This lack of boundedness is an indication that a coupled map lattice of the form (\ref{cml1}) does not provide an appropriate model of a discrete reaction-diffusion system. In fact, a careful analysis shows that a coupled map of the form (\ref{cml1}), for any local dynamics, spatial dimension, and interaction neighborhood, does not have a coherent physical interpretation in terms of a discrete analogue of a reaction-diffusion process~\cite{Roh96}. Systems governed by coupled maps of the form (\ref{cml1}) have the unphysical property that individuals (e.g., molecules or organisms) may disintegrate (or die) and still diffuse (or disperse) (see ~\cite{Roh96} for a detailed discussion). This lack of mathematical well-definedness together with the lack of any coherent physical interpretation has resulted in coupled map lattices with diffusive interactions of the form taken in (\ref{cml1}) falling out of use. 

The solution to the problems alluded to above is simple and elegant --- to separate the reaction (or reproduction) process represented by the local map $f(x)$ from the diffusive process~\cite{Jack90,Roh96}. Each iterate of the coupled map is now viewed as having two stages. In stage 1 (the reaction stage) we have 
\begin{equation}
x_i^\prime (t)=f(x_i(t)).
\label{reaction}
\end{equation}
This is followed by stage 2 (the diffusion stage) to give the final value of the next iterate 
\begin{equation}
x_i(t+1)=x_i^\prime (t)+d[x_{i+1}^\prime(t)+x_{i-1}^\prime(t)-2x_i^\prime(t)].
\label{diffusion}
\end{equation}
Combining these two stages we obtain the complete coupled map lattice 
\begin{equation}
x_i(t+1)=f(x_i(t))+d[f(x_{i+1}(t))+f(x_{i-1}(t))-2f(x_i(t))].
\label{cml2}
\end{equation}
This form of the coupled map lattice generalizes directly to arbitrary spatial dimensions and interaction neighborhoods. This form (\ref{cml2}) of the coupled map lattice completely resolves the two problems associated with the earlier form of the coupled map lattice (\ref{cml1}): it is easy to show that iterates of (\ref{cml2}) are always bounded for reasonable values of $d$ and $x_i (0)$~\cite{Jack90}, and (\ref{cml2}) has a clear physical interpretation as a discrete reaction-diffusion system with a reaction stage followed by a diffusion stage~\cite{Jack90}. 

Due to these advantages all recent work on diffusively coupled map lattices have used the form (\ref{cml2}), and its generalizations. The stability theory of coupled map lattices of the form (\ref{cml2}) has been studied in great generality. It has been shown that for such a map lattice with arbitrary local map $f(x)$, spatial dimension, and interaction neighborhood, if $f(x)$ has a stable fixed point then the corresponding coupled map has a stable homogeneous fixed point~\cite{Roh96,LJans}. Thus, if the local dynamics is trivial it is impossible for any coupled map lattice of the general form of (\ref{cml2}) to produce non-trivial spatial patterns. We note that this general result is in accord with physical intuition: if the local map approaches the same equilibrium value at every site, then it would be most surprising if the introduction of a diffusive spatial coupling (which will act to reduce differences between site variables at different positions) could result in a spatially inhomogeneous equilibrium. It follows from these stability results that non-trivial global dynamics is only possible in any coupled lattice map of the form of (\ref{cml2}) if the local dynamics defined by $f(x)$ is non-trivial, in the sense that it does not have a stable fixed point. It is for this reason that any coupled map of the form (\ref{cml2}) that displays non-trivial global dynamics is  necessarily based on a local map that has non-trivial dynamics.  Examples of such situations include: Ref.~\cite{CMan88}, in which the local dynamics is defined by a chaotic tent map; Ref. ~\cite{CMan92}, where the local dynamics is given by a chaotic logistic map; and Ref.~\cite{Pol93} in which the local map has a stable 3-cycle. 

The notion of coupling local site dynamics via a competitive interaction, as in our model, has naturally arisen in both theoretical ecology
and developmental biology. In the ecological context, the dynamical variable $x_i(t)$ can be interpreted as the size 
of the population at site $i$ and at time $t$.
Since the growth in the population at a given site will be limited both by competition with individuals at that site and also
with individuals from neighboring populations, it is natural to consider competitive coupling in the population dynamics of
such systems. For example, Hara 
and co-workers~\cite{Hara} addressed the effects of local versus global competition on size distributions in tree populations. The site dynamics there
was described by a single logistic differential equation, which exhibits only a stable equilibrium point, which corresponds to a uniform tree size. However, on including
either global or local spatial competition, there emerges a parameter threshold beyond which variation in size distribution is seen.     
A discrete-time model of spatial population dynamics with local
competition has also been addressed in Ref.~\cite{DoebKill}.
The notion of competitive coupling as a basic mechanism for pattern formation is also closely related to the idea of lateral 
inhibition, in which the growth of a structure at a given location inhibits the 
growth of similar structures in a region surrounding the focal object, which has been studied in connection to pattern formation in 
developmental biology~\cite{Lateral}.

It is interesting to note that a similar model to the one we study here has been investigated in Refs.~\cite{N1,N2,N3} in the context of understanding income
distributions. Although the model studied in these papers is formally similar to our model, the aims of our work and that of Refs.~\cite{N1,N2,N3} are very different.
Whereas our focus is on how a homogeneous initial state can undergo spontaneous symmetry breaking and result in stable pattern
formation, Refs.~\cite{N1,N2,N3} do not contain any discussion of spatial pattern formation and are, instead, concerned with modeling wealth or income
distributions in society.

It is also worth mentioning that the form of our model is typical of that of various discrete competition models~\cite{N4,N5}. The novel aspect of our work is that
we show that a simple, coupled map model incorporating spatial competition can produce stable spatial pattern formation, even when the local
dynamics is completely trivial. In contrast, the models studied in ~\cite{N4,N5} are not spatial models and are unconcerned with any aspects of spatial
pattern formation.

Finally, in contrast to the specific (and in some cases rather complicated) models of spatio-temporal dynamics with local 
competition that have been previously investigated, our aim in this paper is to explore in some detail a very simple class of 
discrete-time spatio-temporal dynamical models in which competitive interactions lead to spatial pattern formation. We show that
in this general class of models, pattern formation is a robust consequence of local spatial competition, and does not depend
on the specifics of either the local dynamics or the spatial geometry.

\section{COMPETITIVELY COUPLED MAPS}

We begin, for the sake of generality,  by defining a competitively coupled system of maps 
for an arbitrary network, which we shall refer to as a {\it competitively coupled map network} (CCMN). However,
most of this paper will focus on the case of relevance for pattern formation in which the network is a lattice, 
corresponding to {\it competitively coupled map lattices} (CCML).

We define a CCMN as follows. First, let $f: \mathbb{R} \rightarrow \mathbb{R}$ be a smooth, one-dimensional,
unimodal map. Further, let us assume that $f$ can be written in the form $f(x)=xF(x)$ where $F:\mathbb{R} \rightarrow \mathbb{R}$
is also a smooth map that is monotonically decreasing on the domain of interest. We note that many
celebrated maps $f$ can be expressed in this form. Well-known examples~\cite{May} include the logistic map where
$F(x)=r(1-x)$, the Ricker map where $F(x)=\lambda e^{-ax}$, the Hassell map where $F(x)=\lambda/(1+ax)^b$, and
the Maynard Smith map for which $F(x)=\lambda/(1+ax^b)$. The parameters $r, \lambda$, $a$ and $b$ are positive constants. 
The parameter $a$ can be scaled away so we can, without loss of generality, set $a=1$.
It is well-known that all these maps have qualitatively identical dynamical features,
including stable fixed points, limit cycles and a period-doubling route to chaos~\cite{May}. Here we focus
on pattern formation ariseing from the Ricker, Hassell and Maynard Smith maps, which are a representative sample
of one-dimensional, unimodal maps for which the iterates are always non-negative. We do not consider the logistic map 
as the possibility of the iterates becoming negative results in technical complications which have no bearing on the 
pattern formation mechanism we are investigating. Futhermore, as we wish to emphasize the role competitive interactions play in
pattern formation, we chose to employ maps with the simplest possible dynamical behavior. As such, we always consider here
only parameter values for which the maps have a stable fixed point. We reiterate that for such parameter values, diffusively 
coupled map lattices always evolve to a spatially homogeneous state and pattern formation is impossible. Thus, any pattern formation that 
results in our model is a consequence of local competition and not the complexity of the local dynamics.
 
First we consider in some detail the case in which the local dynamics is given by the Ricker map and then
we will expand the discussion to include the Hassell and Maynard Smith maps.
For the Ricker map, $f(x)=\lambda xe^{-x}$, the fixed points are $\hat{x}=0$ and $x^*= \ln{\lambda}$
given by the condition $F(x^*)=1$. The corresponding stability is determined by the function
$\phi(x)=F(x)+xF'(x)$. A fixed point $x^*$ is linearly stable if $|\phi(x^*)|<1$ and
unstable if $|\phi(x^*)|>1$. For the Ricker map, $\hat{x}=0$ is stable for $0<\lambda<1$ while $x^*$ is
stable for $|1-\ln{\lambda}|<1$, that is for $1<\lambda<e^2$. When $\lambda$ exceeds $1$, the
stable fixed point $\hat{x}=0$ loses stability through a transcritical bifurcation and gives
birth to the stable fixed point $x^*=\ln{\lambda}$. Here we always restrict attention to $\lambda < e^2$ 
corresponding to stable fixed point dynamics.

To form a CCMN based on a map $f(x)=xF(x)$, we first introduce a network $\Gamma$, i.e. $\Gamma$ is a simple,
undirected graph with $m$ vertices (or nodes) which are labeled by $i=1,\cdots, m$. The topology of
$\Gamma$ is determined by the adjacency matrix $A=(a_{ij}), i,j=1,\cdots,m$, where $a_{ij}=1$ if
nodes $i$ and $j$ are connected by an edge (or link) and $a_{ij}=0$ otherwise. Given a node $i\; \in \;\Gamma$
we define the set of neighbors of $i$ to be $N(i)=\{j \; \in \;  \Gamma : a_{ij}=1 \}$. The CCMN associated
with $f$ and $\Gamma$ is a discrete-time dynamical system $\Phi$ on $\mathbb{R} ^m$. The state of this
system at time $t$ is determined by a vector ${\bf x}(t) = (x_1(t),\cdots,x_m(t)) \; \in \; {\mathbb{R}}^m$,
where $x_i(t)$ represents the value of the state variable associated to node $i$ of the network at time $t$.
The dynamical system $\Phi$ gives the time evolution of the state ${\bf x}(t)$, i.e. $\Phi({\bf x}(t))={\bf x}(t+1)$, 
where $\Phi$ is defined by:
\begin{equation}
x_i(t+1)=x_i(t)F\left[ x_i(t) + \sum_{j\; \in \; N(i)} \alpha x_j(t) \right].
\label{eqn2}
\end{equation}
for all $i \in \Gamma$. In this expression $\alpha$ is a non-negative parameter that represents the strength of the 
competitive interaction between site $i$ and a neighboring site $j$.

The first step in understanding the dynamics of (\ref{eqn2}) is to study its fixed points and their stability. In discussing
the fixed points of $\Phi$ it is convenient to assume that the network $\Gamma$ is a regular graph of degree $k$, which
means that every node has exactly $k$ edges incident upon it. In this case, there are two homogeneous fixed points of (\ref{eqn2}), where
this term refers to every node $i\; \in \; \Gamma$ having the same value for the state variable. The first of these is the trivial 
fixed point ${\bf \hat{x}}= {\bf 0}=(0,0,\cdots,0)$ while the non-trivial fixed point is ${\bf x^*}=(x^*,x^*,\cdots,x^*)$, where $x^*$
satisfies $F[(1+k\alpha)x^*]=1$. For the case of the Ricker map, this latter fixed point is given by $x^*=\frac{\ln{\lambda}}{1+k\alpha}$.

The stability of the fixed points is determined by the eigenvalues of the Jacobian matrix $J({\bf x})$ which has elements
$J_{ij}({\bf x}) = \frac{\partial \Phi_i(x)}{\partial x_j}$. Direct calculation from (\ref{eqn2}) shows that the Jacobian evaluated
at the trivial fixed point ${\bf \hat{x}}={\bf 0}$ is $J({\bf 0})=F(0)\cdot I$, where $I$ is the $m \times m$ identity matrix. Thus, every
eigenvalue is equal to $F(0)$, which means simply {\it that the eigenvalue of the original uncoupled map determines the stability
at the fixed point} ${\bf \hat{x}}={\bf 0}$. Therefore, the CCMN dynamics is trivial if and only if the dynamics of the uncoupled map $f$
is stable for $x=0$, which for the Ricker map means $\lambda <1$.

Considering now the non-trivial, homogeneous fixed point ${\bf x^*}$, we find that the Jacobian evaluated at ${\bf x^*}$ is
\begin{equation}
J({\bf x^*}) = (1+x^*F'[(1+k\alpha)x^*])\cdot I + (\alpha x^* F'[(1+k\alpha)x^*])\cdot A \;,
\label{eqn3}
\end{equation}
where $A$ is the $(m \times m)$ adjacency matrix of $\Gamma$. The stability of ${\bf x^*}$ is determined by the $m$ eigenvalues $\psi_j$
of $J$. 

It follows immediately from eq.(\ref{eqn3}) that $J({\bf x^*})$ is a circulant matrix whenever $A$ is a circulant. Note
that circulant matrices are fully specified by the first row vector as all other rows are cyclic permutations of the first row
where the offset corresponds to the row index~\cite{Davis}.
Those networks for which the adjacency matrix is a circulant form an interesting class, known as circulant graphs. 
The networks we study here are all either circulant graphs or are networks that can be directly related to circulant 
graphs. For such networks, the eigenvalues $\psi_j$ of $J({\bf x^*})$ can be directly computed using (\ref{eqn3}).

For the Ricker map, the non-trivial fixed point is $x^*=\frac{\ln{\lambda}}{1+k\alpha}$ and the network Jacobian (\ref{eqn3}) evaluated
at this fixed point is
\begin{equation}
J({\bf x^*})= \left( 1 - \frac{\ln{\lambda}}{k\alpha+1} \right)\cdot I - \left( \frac{\alpha\ln{\lambda}}{k\alpha+1} \right)\cdot A. 
\label{eqn4}
\end{equation}
Given any circulant graph $\Gamma$ it is possible to explicitly compute the eigenvalues $\psi_j, j=0,\cdots,m-1$, of this Jacobian from
which the stability of the non-trivial fixed point in the CCMN can be established.

\section{SPATIAL PATTERN FORMATION}

Within the general framework of a CCMN, we will now focus on such systems in the cases of most interest for pattern formation, when the network $\Gamma$ is a 
spatial lattice in one or two dimensions. We begin with the simpler case where $\Gamma$ is a one-dimensional
lattice with $m$ nodes subject to periodic boundary conditions. Let us also restrict the analysis to nearest neighbor interactions
where site $i$ interacts with sites $(i-1)$ and $(i+1)$. The generalization to other interactions is straightforward. 
Thus, in this case, $k=2$ and the homogeneous fixed point of the competitively coupled 
map lattice is $x^*=\frac{\ln{\lambda}}{2\alpha+1}$. Given the periodic boundary conditions, the Jacobian (\ref{eqn3}) is a symmetric circulant matrix
and the associated real eigenvalues can be explicitly computed from the properties of circulant matrices~\cite{Davis}. This calculation
gives $\psi_0=\ln{\lambda}=\phi(x^*)$, which is the eigenvalue of the uncoupled Ricker map evaluated at the non-trivial fixed point $x^*=\ln{\lambda}$
of the map. Thus, $|\psi_o| < 1$ whenever the fixed point $x^*=\ln{\lambda}$ of the Ricker map is stable. However, if $x^*=\ln{\lambda}$ is unstable for the 
uncoupled Ricker map, then $|\psi_0| > 1$ and the homogeneous fixed point $x^*=\frac{\ln{\lambda}}{2\alpha+1}$ of the CCML is also unstable.

Consequently, stability of the fixed point $x^*=\ln{\lambda}$ is a necessary condition for the stability of the homogeneous fixed point ${\bf x^*}$ of the CCML, but 
it is not sufficient. This may be seen by computing $\psi_{m/2}$, for $m$ even, or $\psi_{(m-1)/2}$, for odd $m$. For even $m$ we obtain 
$\psi_{m/2} = 1 + \frac{2\alpha-1}{2\alpha+1} \ln{\lambda}$, which is the same result as that for the odd $m$ case, $\psi_{(m-1)/2}$, for $ m >> 1$.
It is straightforward to show from the eigenvalues of $J({\bf x^*})$ that if $|\psi_0| < 1$, (i.e. for $1< \lambda < e^2$), then the stability of the
homogeneous fixed point ${\bf x^*}$ is determined by $\hat{\psi} = 1+\frac{2\alpha-1}{2\alpha+1} \ln{\lambda}$, for $m >>1$. Thus, ${\bf x^*}$ is
stable if $|\hat{\psi}| < 1$, i.e. for $\alpha < 1/2$ and unstable for $\alpha > 1/2$. {\it This clearly demonstrates that the competitive coupling
results in the stability of the non-trivial homogeneous fixed point being distinct from the stability of the nodal Ricker map dynamics.Therefore, for
$1 < \lambda < e^2$ (i.e. for trivial site dynamics), $\alpha > 1/2$ results in spontaneous symmmetry breaking of the spatially homogeneous initial state.} 
As we shall show, this results in the emergence of complicated stable spatial patterns.

In this one dimensional situation, the dynamics at a site $i$ given by (\ref{eqn2}) takes the form
\begin{equation}
x_i(t+1) = \lambda x_i(t) e^{-\left[ x_i(t) + \alpha(x_{i-1}(t)+x_{i+1}(t)) \right]} \;.
\label{eqn5}
\end{equation}
It is apparent from (\ref{eqn5}) that our
CCML naturally incorporates the principle of local activation and longer-ranged inhibition that is the essential, 
but indirectly realized, mechanism of pattern formation
in Turing-type systems~\cite{Meinh}.
It is also clear from the negative exponential in (\ref{eqn5}) that a site surrounded by neighbors with large values of the 
site variable $x$ will feel strongly the effects of the coupling, 
and will be suppressed.  However, while a strong coupling has an inhibitory effect on a site between two neighbors with 
large $x$ values, the same coupling can 
leave the larger neighbors relatively unaffected, since low surrounding sites have little inhibitory effect on their neighbors. 
 This is in strong contrast to
diffusive coupling~\cite{Kaneko}, which for a single species can only act to uniformize the levels of the site variable~\cite{WallKap,LJans}. It is this difference which is responsible
for the pattern formation seen in our model.

\begin{figure}[htb]
\centering{
\subfigure[]{
\includegraphics[scale=.29]{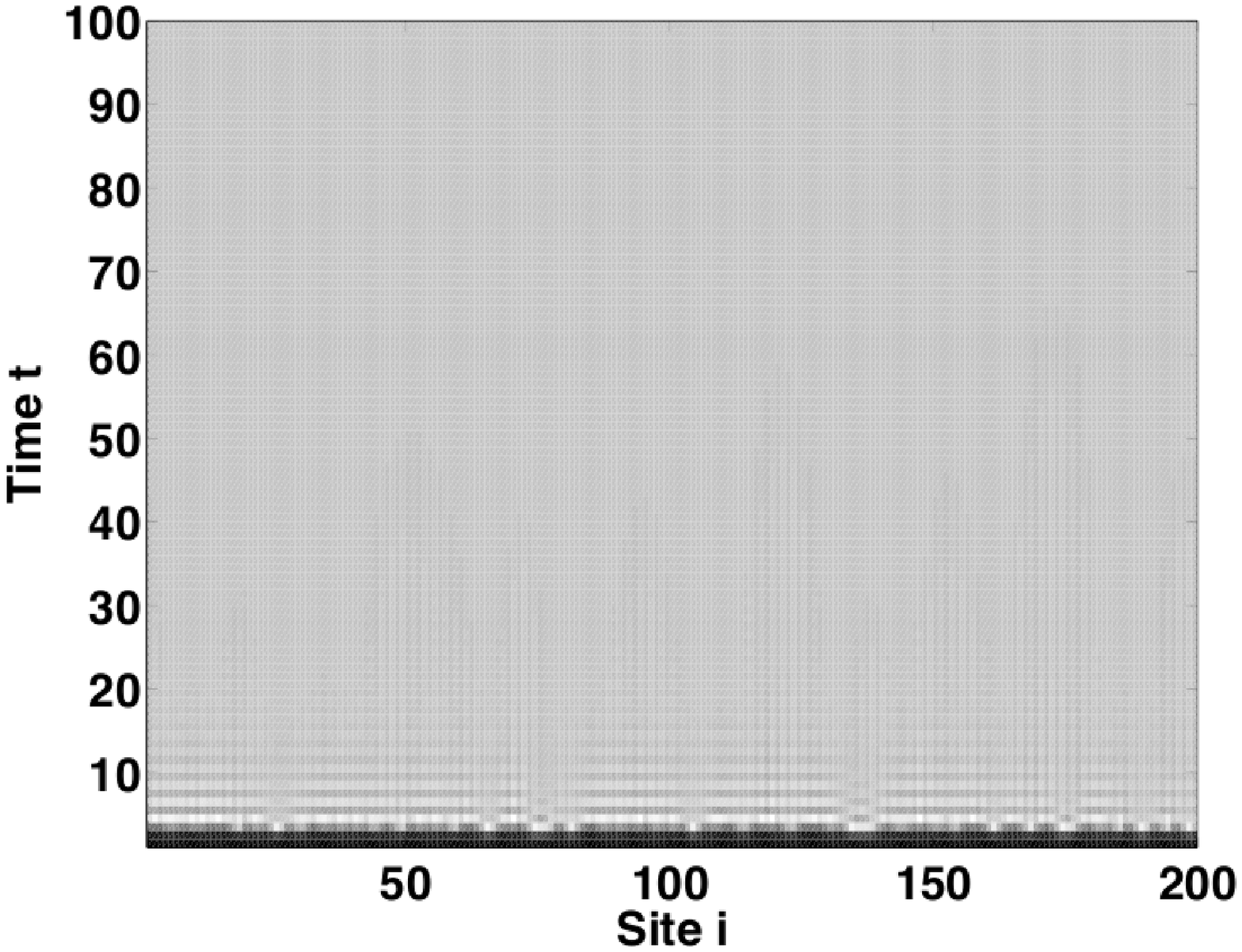}
\label{Fig1a}}
\subfigure[]{
\includegraphics[scale=.29]{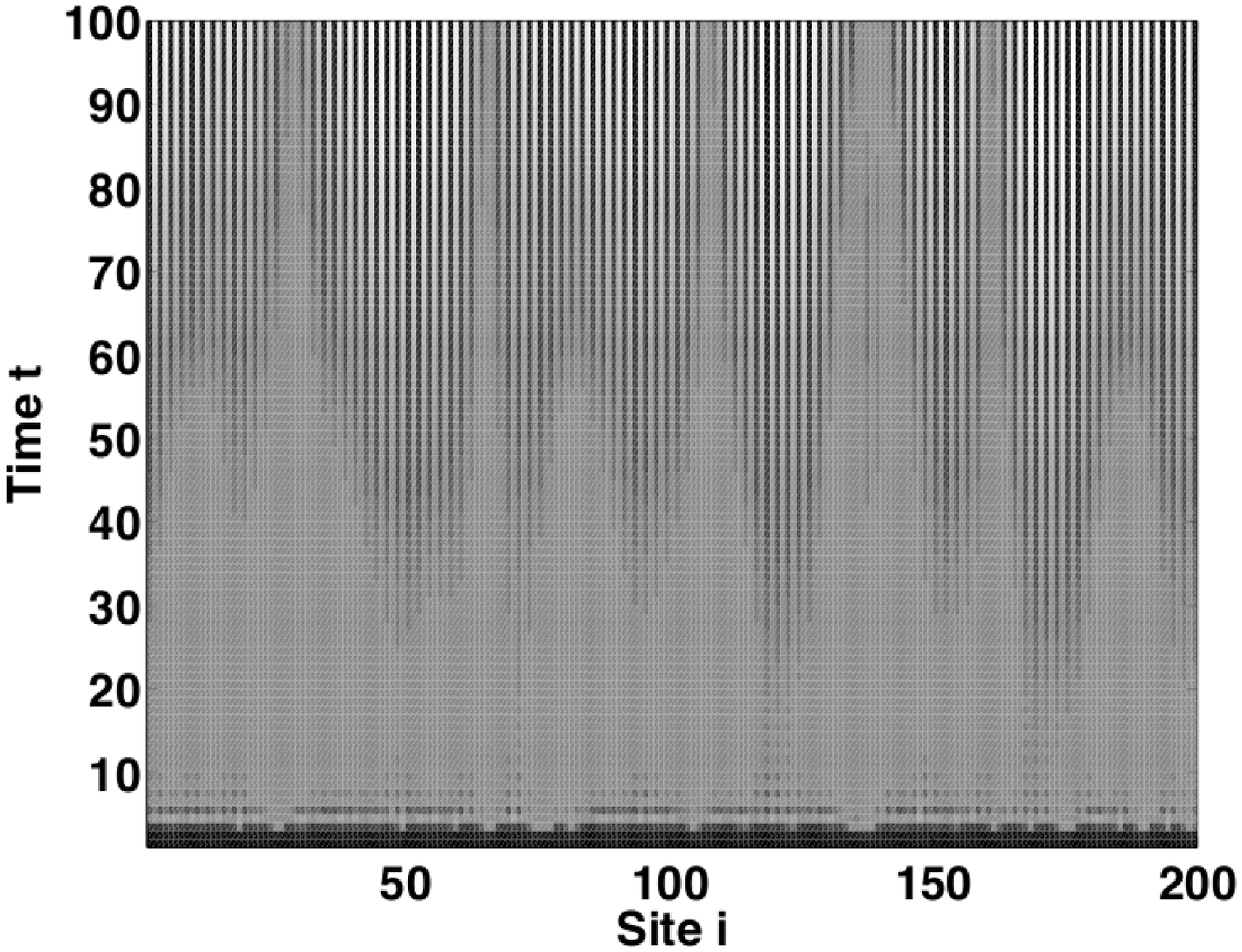}
\label{Fig1b}}
}
\vspace{-0.1in}
\caption{Evolution of the population with time $t$ on a one-dimensional chain with $200$ nodes. In (a), the coupling
$\alpha=0.49$ which is smaller than the threshold value. Panel (b) corresponds to $\alpha=0.53$ and
results in a non-trivial pattern. In both cases, the parameter $\lambda=6$ for which the site dynamics has a 
stable fixed point.}
\label{Fig1}
\vspace{-0.15in}
\end{figure}

Figure \ref{Fig1} illustrates the change in stability in our model at $\alpha=1/2$ for a one-dimensional chain with periodic boundary conditions. In Fig.~\ref{Fig1a}, the differences
in initial populations at the sites are quickly erased and the stable, non-trivial, spatially homogeneous state is reached. By contrast, on crossing the threshold $\alpha=1/2$,
a very interesting spatial pattern emerges as seen in Fig.~\ref{Fig1b}. We note that at early times, the alternating pattern discussed earlier is clearly visible which then
evolves to a more complex pattern in time. This transition happens progressively later in time as one approaches the $\alpha=1/2$ threshold. It should be emphasized that for this value
of the map parameter $\lambda=6 < e^2$, the uncoupled site dynamics would converge to a fixed point which clearly demonstrates that the complexity seen in the spatial pattern is
an effect of symmetry breaking.

Let us now consider $\Gamma$ to be a two-dimensional square lattice with nearest neighbor interactions, where each site
$(i,j)$ is coupled with sites $(i-1,j), (i,j-1), (i,j+1)$ and $(i+1,j)$. The
map dynamics is given by the appropriate specialization of (\ref{eqn2}). 
Our earlier analysis can be readily extended to this
situation to show that $\alpha=1/4$ is the critical value of the inter-site coupling. 

For $\lambda < e^2$, the Ricker map evolves to a stable fixed point, $x^*=\ln{\lambda}$. For these $\lambda$ values and with coupling
$\alpha < 1/4$, the homogeneous fixed point $x^*$ of the CCML is stable and the value at each site converges to $\ln{\lambda}/(1+4\alpha)$.
When the coupling reaches $\alpha = 1/4$ we see the first qualitative change in the dynamics of the CCML, a ``checkerboard'' pattern consisting of 
alternating high and low values. The checkerboard pattern arises as a consequence of the short range 
activation and long-range inhibition inherent in competitive coupling. 

\begin{figure}[htbp]
\centering{
\hspace{-0.35in}
\subfigure[]{
\includegraphics[scale=.308]{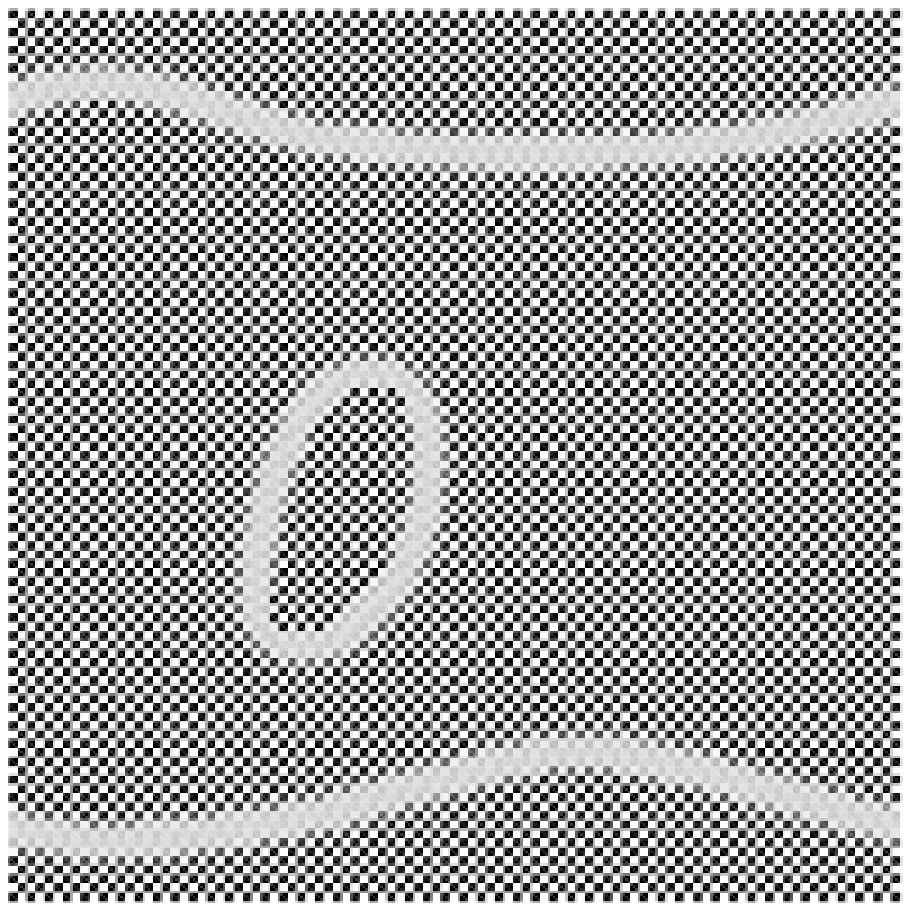}
\label{Fig2a}}
\vspace{-0.199in}
\hspace{-0.25in}
\subfigure[]{
\includegraphics[scale=.379]{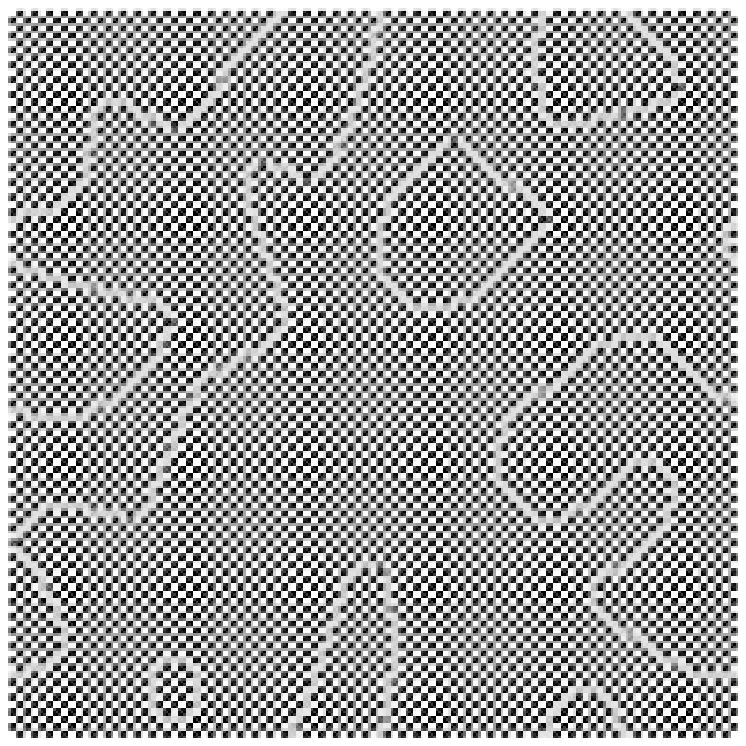}
\label{Fig2b}}
}
\caption{Typical patterns seen in two-dimensional, nearest neighbor CCMLs, illustrated for the Ricker map. In both cases, $\lambda=5, N=100^2$, (a) $\alpha=0.26$ and
shows both toroidal patterns and a long-term transient loop.
 (b) $\alpha=0.35$ }
\label{fig2}
\vspace{-0.08in}
\end{figure}

At $\alpha=1/4$ the differences in neighboring sites evolves slowly, and the checkerboard emerges after several tens of thousands of iterations.  
However, with just slightly larger coupling ($\alpha=0.26$), the formation of the checkerboard becomes much more rapid.  In particular, the 
checkerboard pattern forms at several different regions of the lattice at the same time. This leads to the formation of ``domain walls'' between 
neighboring checkerboard patterns when the high and low sites are out of phase. As seen in Figs. \ref{Fig2a}, the boundaries of 
these domains either wrap around the toroidal geometry or form simple closed loops on 
the lattice. At these values of the coupling, simple loops on the lattice are transient patterns which ultimately shrink to a point, over times which are
larger than $10^5$ iterations. By contrast, 
the toroidal patterns are stable and may be viewed as topological defects associated with the 
non-trivial first homotopy group of the torus. As $\alpha$ increases, the minimum and maximum values of the underlying 
checkerboard pattern do not vary significantly.  However, the increased coupling strength results in the checkerboard emerging more rapidly and at more points 
on the lattice, resulting in complicated domain wall structures. Additionally, with the stronger coupling, isolated closed loops no longer decay but 
become as persistent as the toroidal patterns.  (Fig.~\ref{Fig2b}). 

Having illustrated the ability to generate stable spatial patterns, we now include the effects of next-nearest 
neighbor interactions, where site $(i,j)$ interacts with all eight
sites surrounding it. The threshold value of $\alpha$
can be verified to remain at $1/4$. We observe that in the next-nearest neighbor CCML the checkerboard pattern is 
no longer a feature of the model.  While the local activation and long range inhibition still persist, next 
nearest neighbor interactions allow or a new possibility, namely, the occurence of geometrical frustration~\cite{Sadoc}. The resulting 
structure closely resembles glassy spin systems~\cite{SG}. 

\begin{figure}[htbp]
\centering{
\vspace{-0.07in}
\hspace{-0.6in}
\subfigure[]{
\includegraphics[scale=.15]{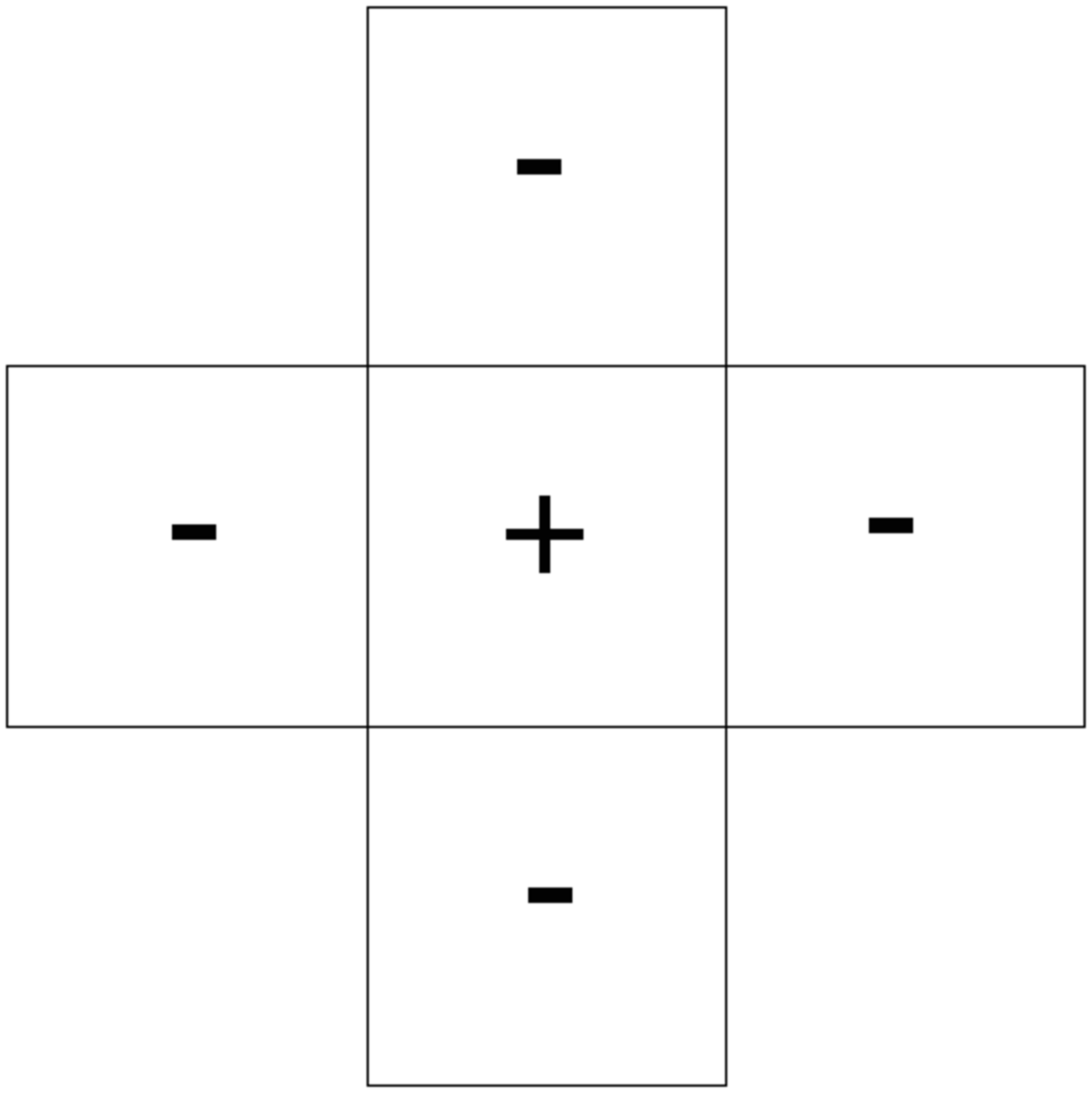}
\label{Fignn}} 
\hspace{-0.5in}
\subfigure[]{
\vspace{0.4in}
\includegraphics[scale=.15]{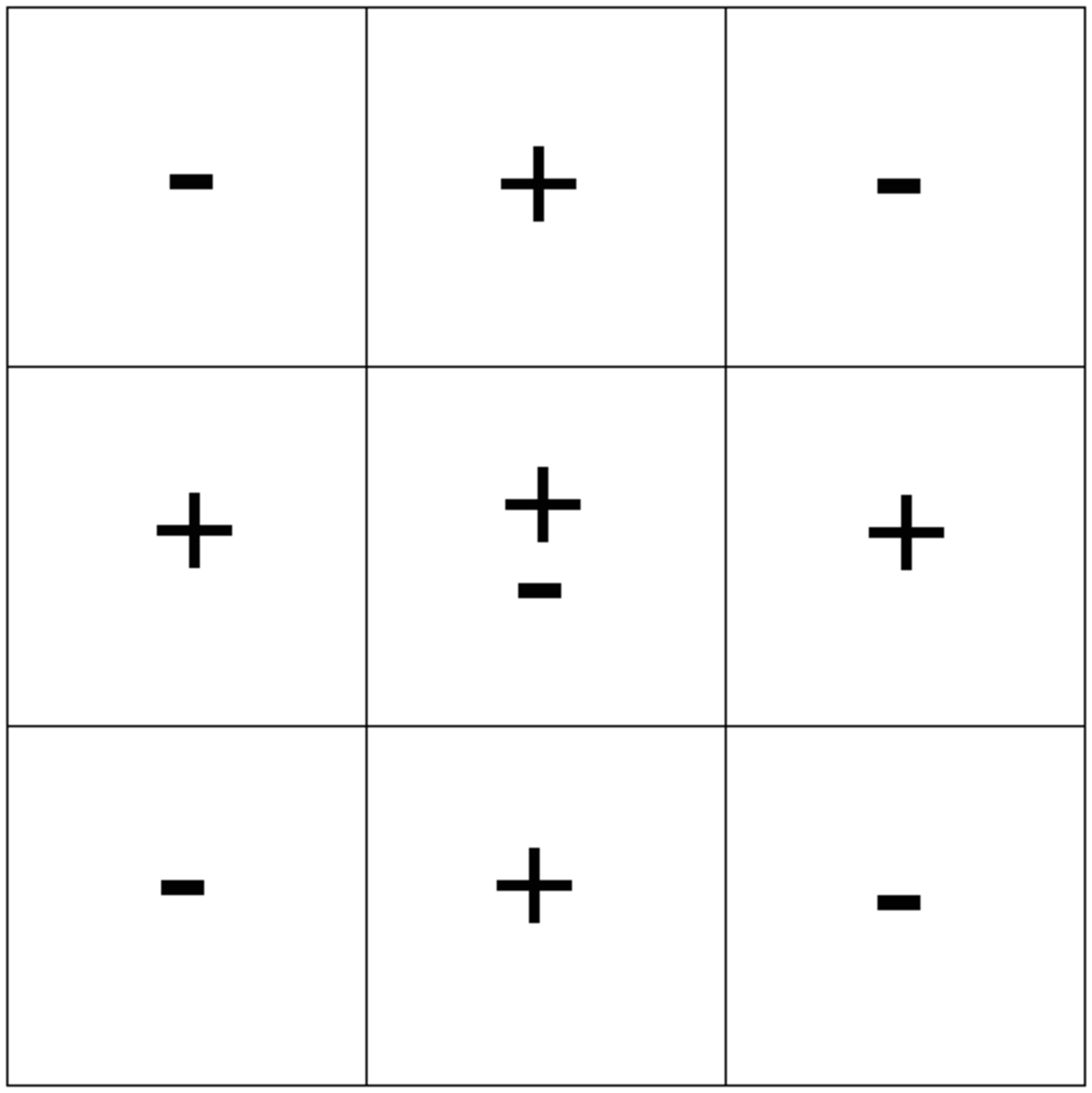}
\label{Fignnn}}
}
\caption{Schematics illustrating (a) the basic checkerboard pattern arising with nearest-neighbor
interactions and (b) the disappearance of the effect due to geometrical frustration once next-nearest 
neighbor interactions are included. The $+$ and $-$ symbols refer to high and low site variable values, respectively.
The symbol $\pm$ in the central cell of (b) indicates that, with next-nearest neighbor interactions, it is not
possible for all nine site variables in the fundamental $3 \times 3$ neighborhood to assume their dynamically
preferred values. This obstruction is the origin of the geometrical frustration that occurs in CCMLs with
next-nearest neighbor interactions.}
\label{FigFrus}
\vspace{-0.05in}
\end{figure}

\begin{figure}[htbp]
\centering{
\vspace{-0.07in}
\hspace{-0.3in}
\subfigure[]{
\includegraphics[scale=.32]{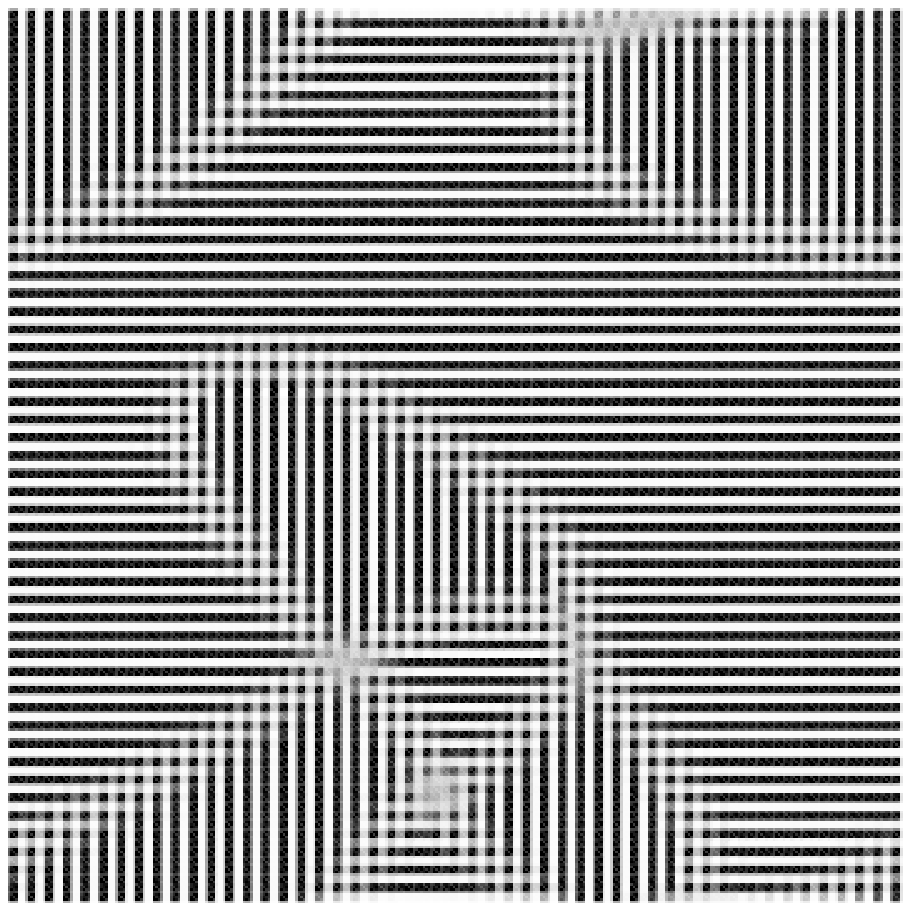}
\label{Fig3a}} 
\hspace{-0.5in}
\subfigure[]{
\vspace{0.4in}
\includegraphics[scale=.255]{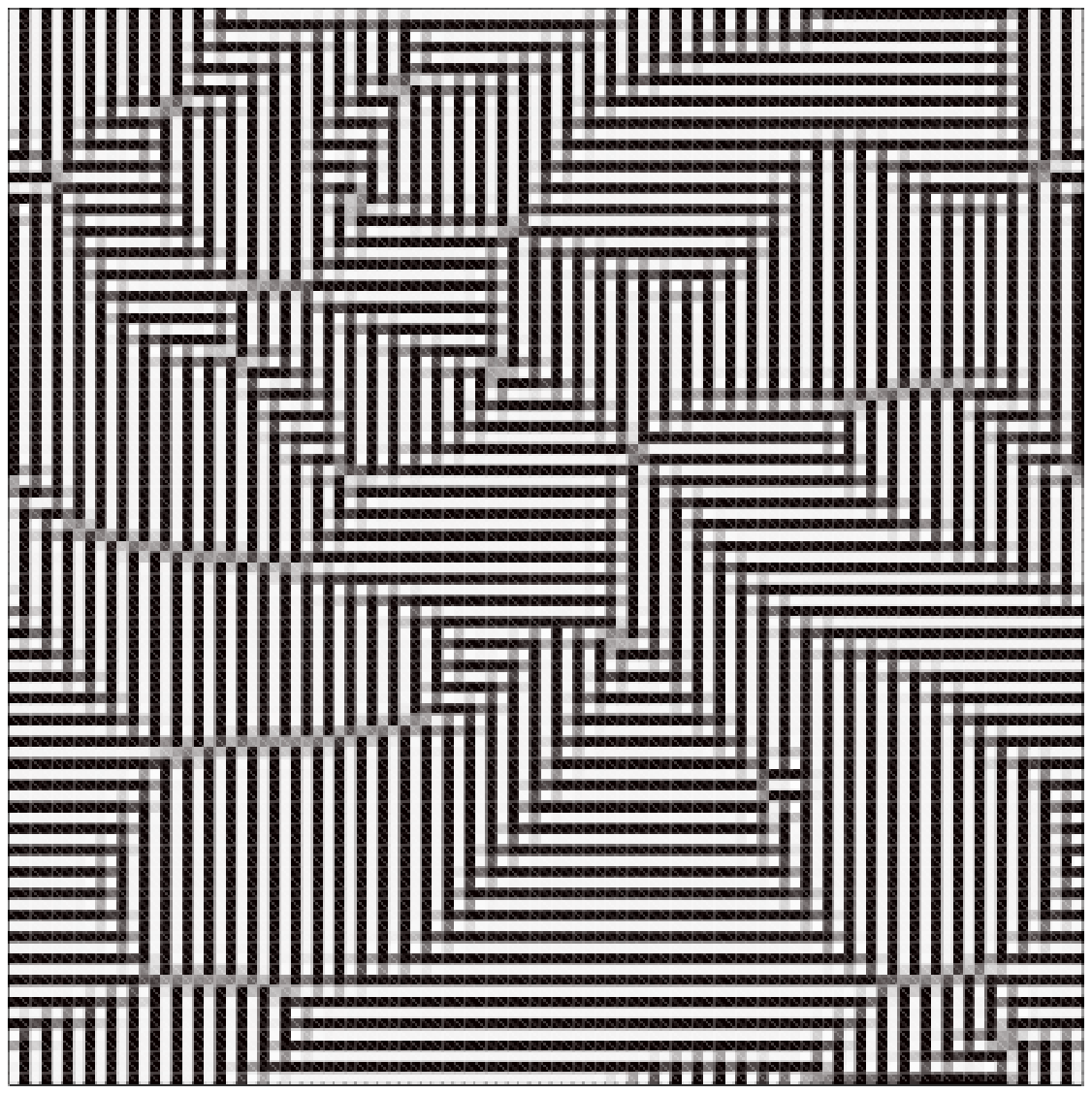}
\label{Fig3b}}
}
\caption{Typical patterns in next-nearest-neighbor CCMLs. $\lambda=5$
$N=100^2$ and (a) $\alpha=0.26$ and (b) $\alpha=0.35$.}
\label{Fig3}
\vspace{-0.05in}
\end{figure}

Local competitive interactions tend to result in sites with high values of
the dynamical variable having neighbours with low values, and vice versa. With nearest-neighbor interactions there is no geometric
obstruction to this occuring as shown in Fig.~\ref{Fignn}. However, next-nearest-neighbor interactions introduce a geometrical
obstruction to such a distribution of dynamical site values, as shown in Fig.~\ref{Fignnn}. The occurrence of this geometrical frustration results
in the appearance of domains with distinct orientation as shown in Fig~\ref{Fig3}. 
As seen from the second of the two panels shown, the domains get smaller with increasing $\alpha$. We note that 
in all the cases shown, qualitative aspects of the results are independent
of the initial conditions. Specifically, even very small (one part in $10^6$) initial spatial variations evolve 
dynamically into the spatial patterns shown.
 
\begin{figure}[htbp]
\centering{
\vspace{0.07in}
\hspace{-0.33in}
\subfigure[]{
\includegraphics[scale=.245]{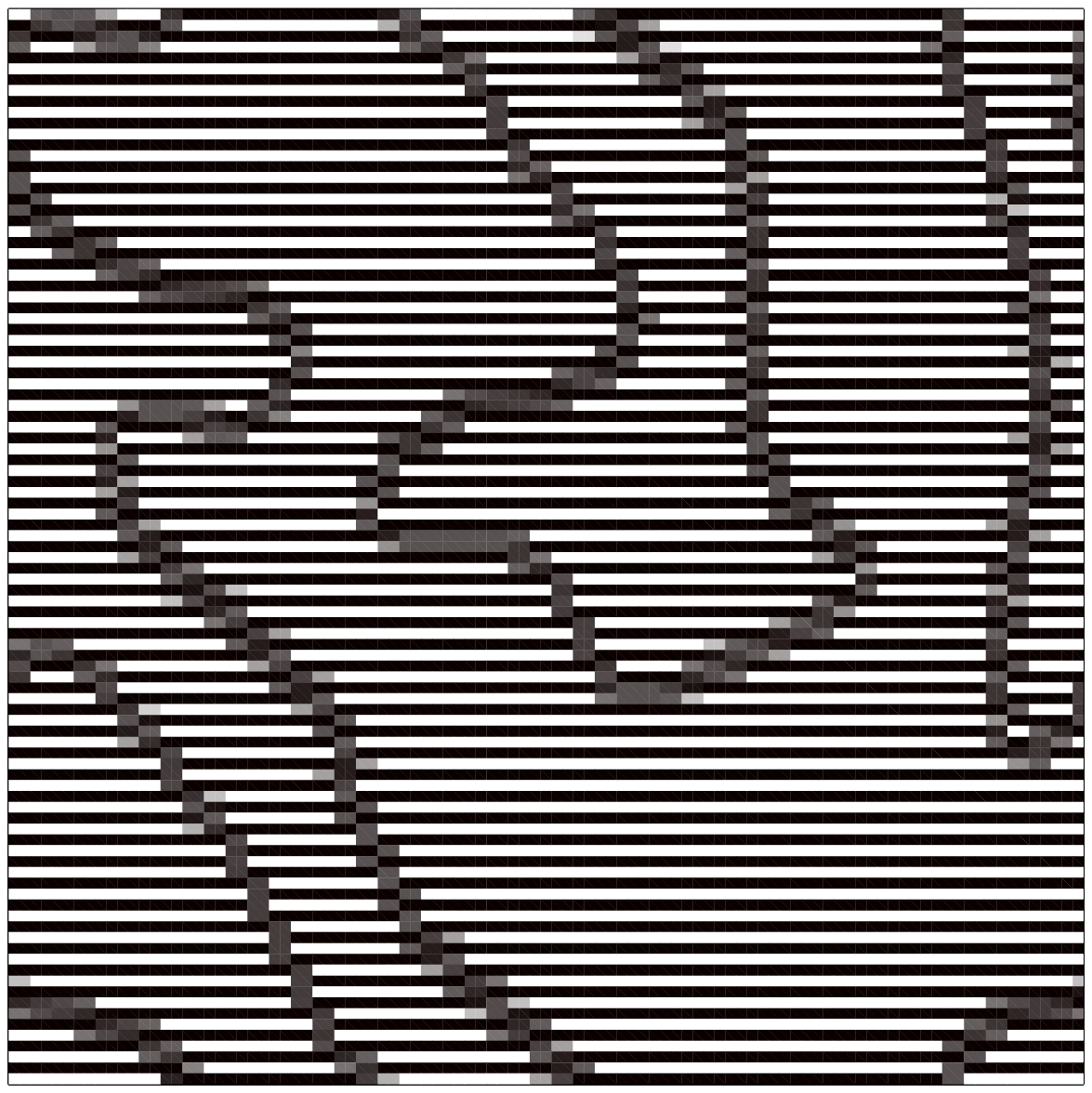}
\label{Fig4a}} 
\hspace{-0.58in}
\subfigure[]{
\vspace{0.4in}
\includegraphics[scale=.255]{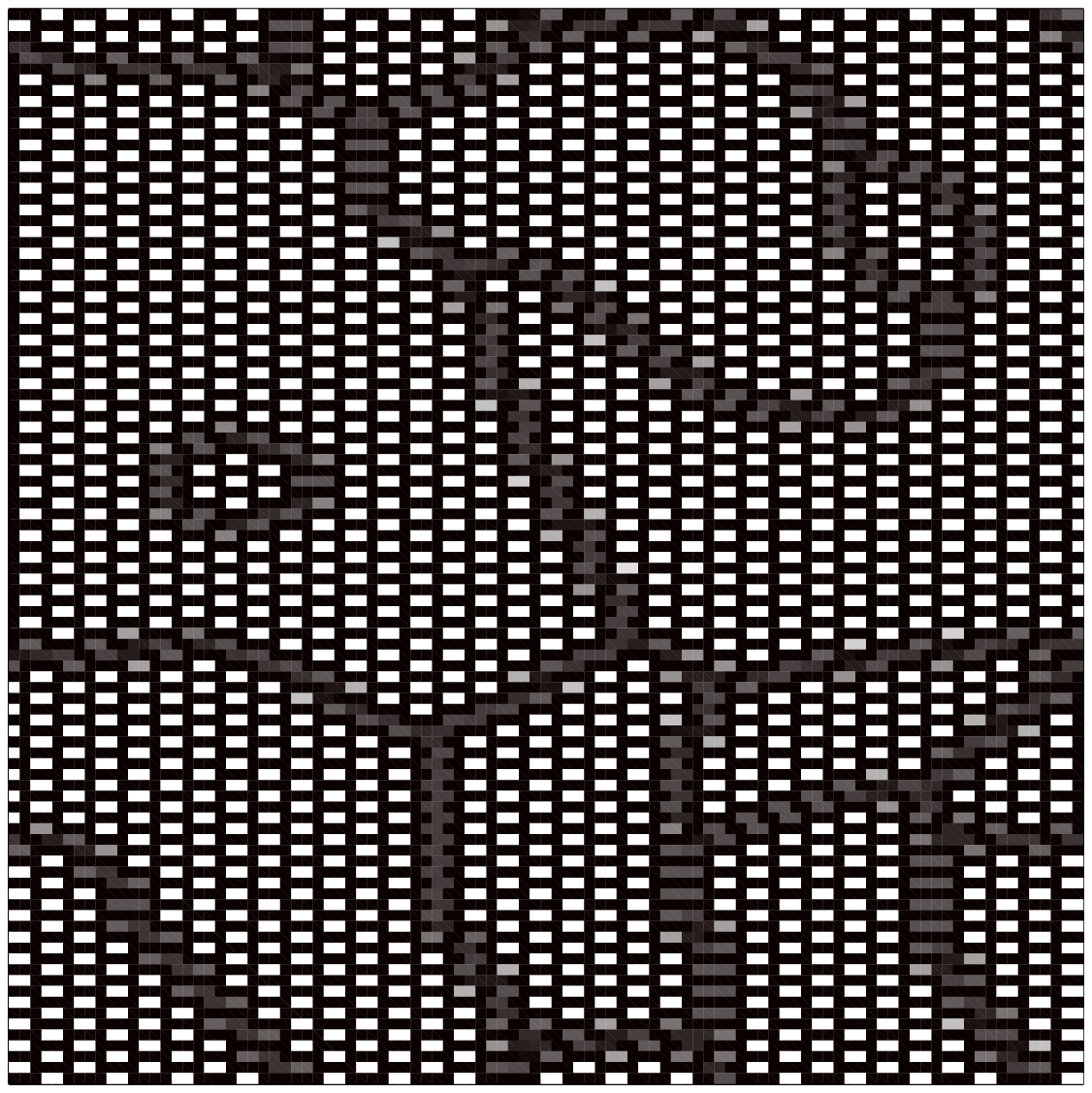}
\label{Fig4b}}
}
\caption{Typical patterns in nearest-neighbor CCMLs on (a) hexagonal and (b) triangular lattices. $\lambda=5$,
$N=100^2$ and $\alpha=0.45$ in both cases. In each of these situations, the critical value of $\alpha$ is $1/3$.}
\label{Fig4}
\vspace{-0.05in}
\end{figure}

The pattern formation mechanism based on competitively coupled maps is applicable to any spatial geometry. In Fig.~\ref{Fig4} we show
typical spatial petterns that result on triangular and hexagonal two-dimensional lattices, with nearest neighbor interactions and
local Ricker map dynamics. The pattern formation that occurs in these lattices is qualititatively very similar to that which takes
place on square lattices with nearest-neighbor interactions. This illustrates the important result that pattern formation in our system 
results from the competitive interaction and is not dependent on the details of the spatial geometry.

Most of our discussion of pattern formation thus far has focused on local dynamics defined by the Ricker map. However, as we now show,
pattern formation occurs in CCML when the local dynamics is described by any unimodal map. We illustrate this by considering 
local dynamics defined either by the Hassell map, $f(x)=\lambda x/(1+ax)^b$, or the Maynard Smith map, 
$f(x)  = \lambda x/(1+ax^b)$~\cite{May}. Here again, we restrict ourselves to parameter values for which these maps have only stable fixed
points so as to make clear that any pattern formation is an outcome of local, competitive, spatial interaction and not a result
of any dynamic complexity of the local map.

\begin{figure*}[bht]
\vspace{-0.11in}
\centering{
\subfigure[]{
\includegraphics[scale=.38]{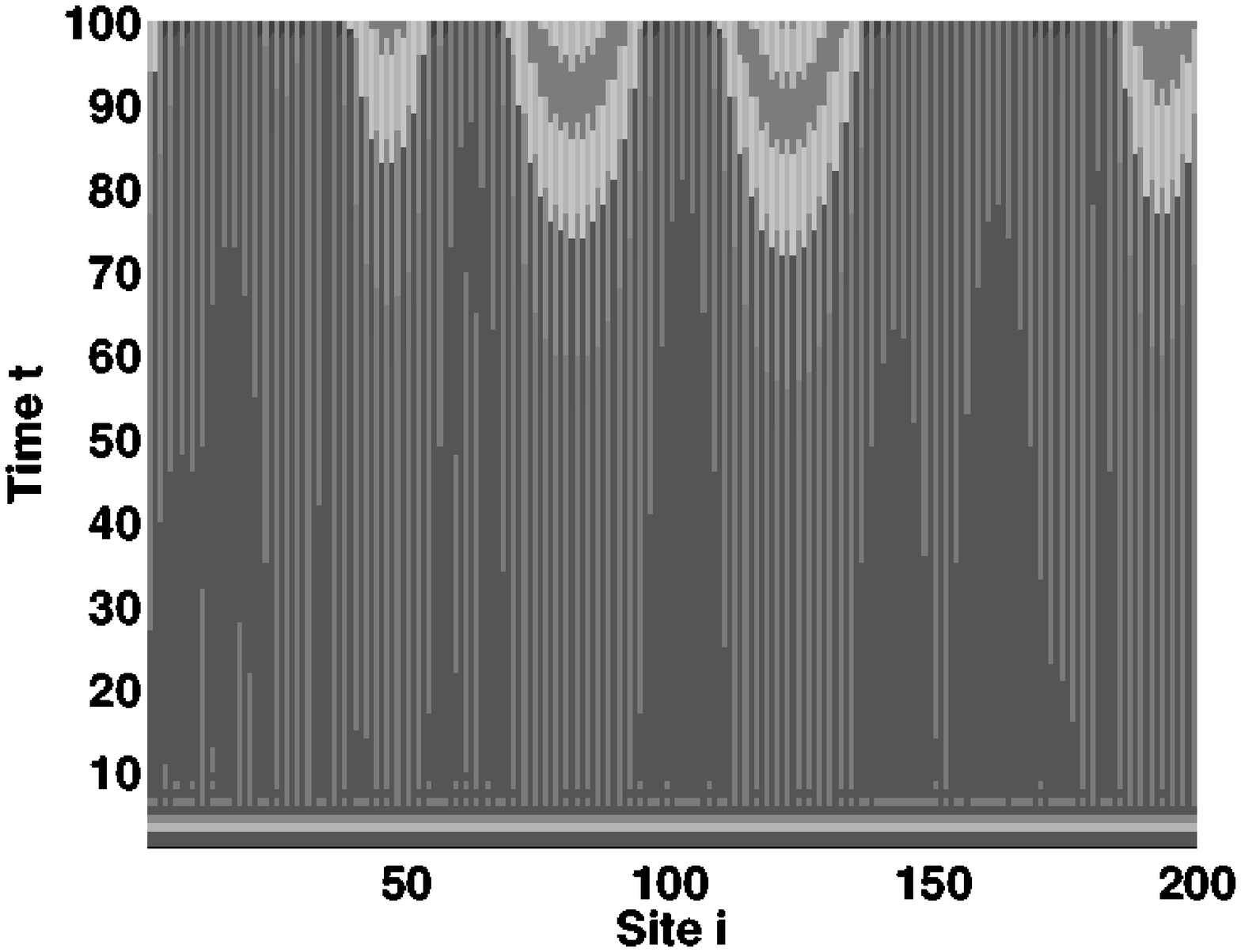}
\label{Fig5a}}
\vspace{-0.5in}
\subfigure[]{
\includegraphics[scale=.38]{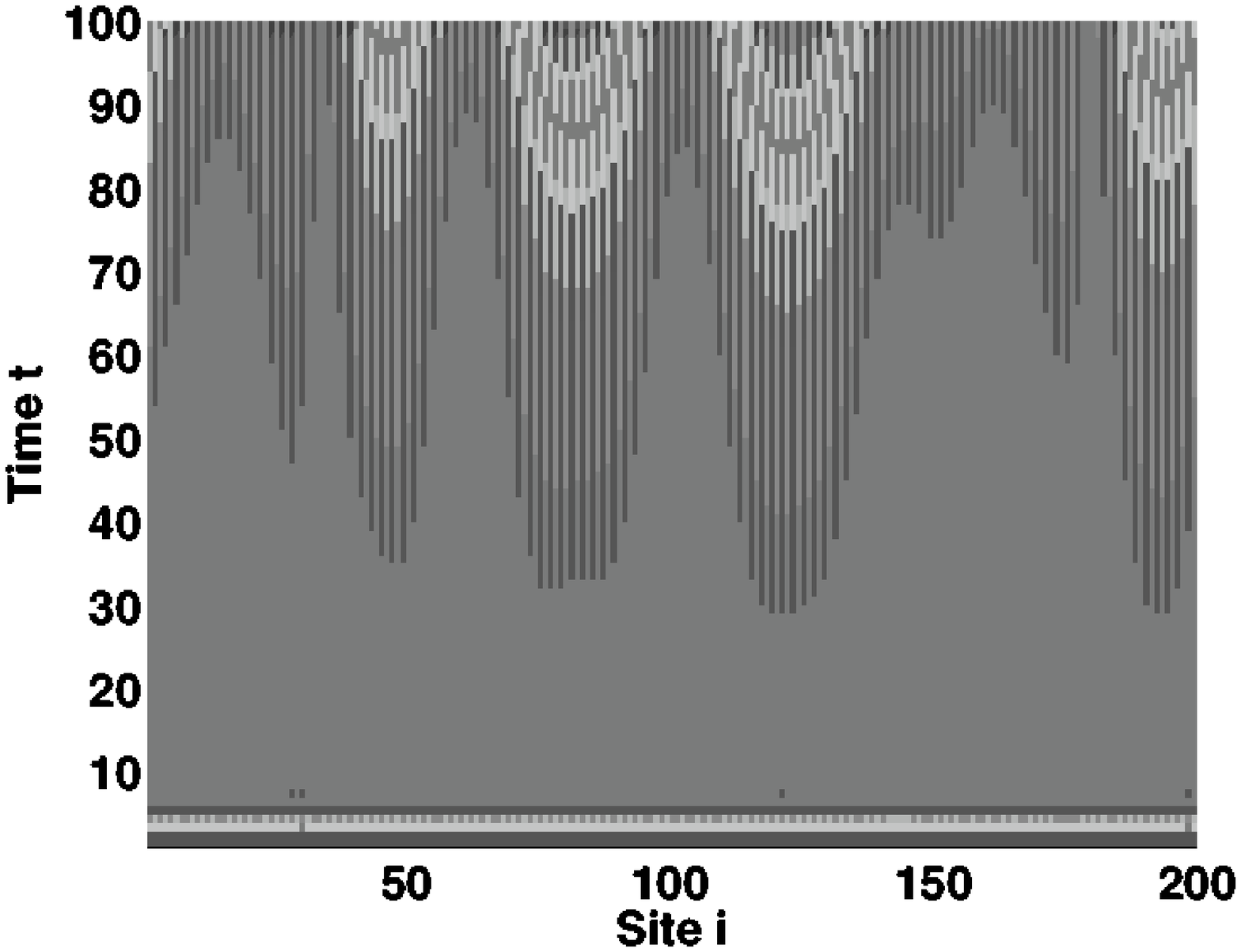}
\label{Fig5b}}
}
\vspace{0.32in}
\caption{Evolution of the population with time $t$ on a one-dimensional periodic chain with $200$ nodes. In (a), the Hassell 
map governs the local dynamics with $b=2$ and $\lambda=9$ while panel (b) uses the Maynard Smith map for parameter values
of $b=2$ and $\lambda=4$. The inter-site coupling $\alpha=0.53$ in both cases which is the same value used for the Ricker
map case in Fig.~\ref{Fig1} and it results in similar non-trivial patterns.}
\label{Fig5}
\vspace{-0.21in}
\end{figure*}

The stability analysis of the CCML with Hassell or Maynard Smith maps is fully analogous to the Ricker map case discussed. 
Thus, it may be shown that, for a one-dimensional lattice with nearest-neighbor interactions, the critical value of $\alpha$ remains
$\alpha=1/2$. Similarly, the critical value for both Hassell and Maynard Smith maps on a two-dimensional square lattice with nearest-
neighbor interactions is $\alpha=1/4$. We show in Fig.~\ref{Fig5} pattern formation for CCML with Hassell and Maynard Smith local
dynamics in a one-dimensional latice geometry. The results are qualitatively similar to those seen earlier with Ricker map local
dynamics. 

\begin{figure}[thb]
\centering{
\vspace{0.04in}
\hspace{-0.38in}
\subfigure[]{
\includegraphics[scale=.19]{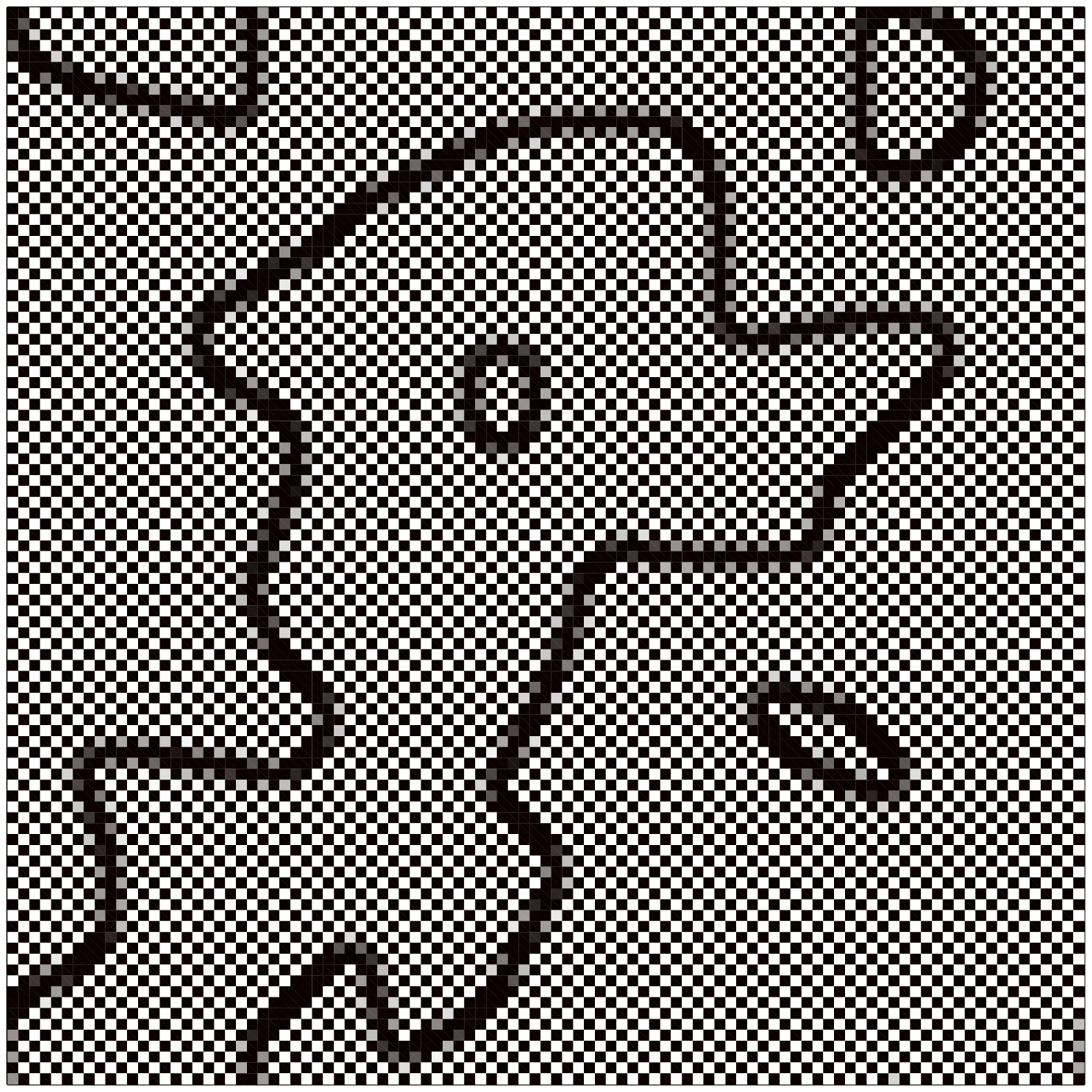}
\label{Fig6a}} 
\hspace{-0.68in}
\subfigure[]{
\vspace{0.39in}
\includegraphics[scale=.185]{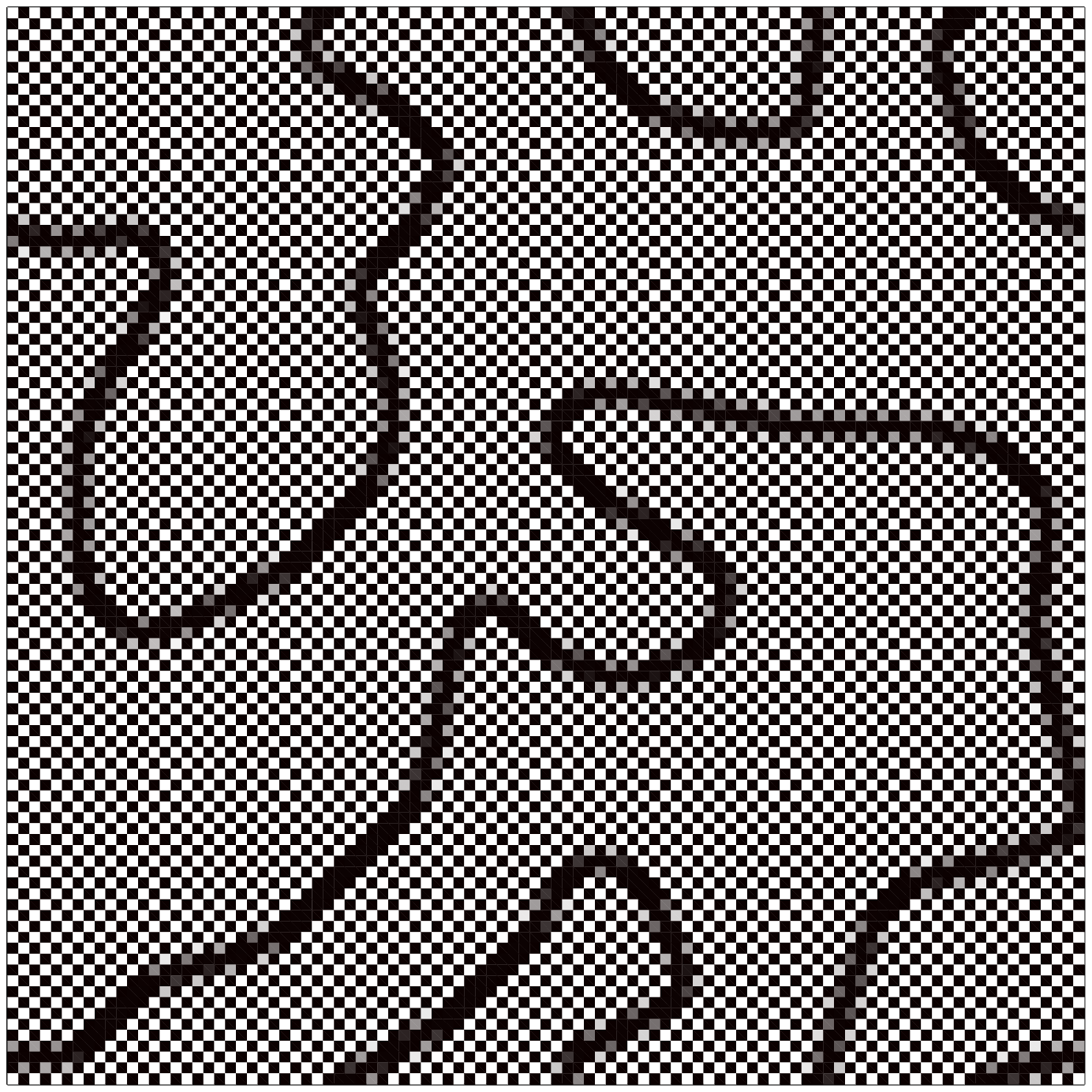}
\label{Fig6b}}
}
\vspace{-0.08in}
\caption{Typical patterns in nearest-neighbor CCMLs with (a) Hassell and (b) Maynard-Smith local map dynamics. 
The relevant parameter values are (a) $b=2$ and $\lambda=9$ and (b) $b=2$ and $\lambda=4$.
A square lattice with $N=100^2$ sites and $\alpha=0.26$ was considered in both cases. }
\label{Fig6}
\vspace{-0.05in}
\end{figure}

In Fig.~\ref{Fig6}, pattern formation for the case of a square lattice with Hassell and Maynard Smith local dynamics
are shown with nearest-neighbor interactions, which are again very similar to those seen earlier for the Ricker map dynamics. Adding
next nearest neighbors results once again in the appearance of geometric frustration in the dynamics. These results make clear that pattern
formation in CCML is a robust property which results from local competitive interactions and not the specifics of either the spatial
geometry or the local dynamics.

\section{CONCLUSIONS}

We have studied here a new  model of spatial pattern formation based on a novel type of map lattice, in which the fundamental spatial
interactions are competitive, rather than the diffusive ones that are typically considered. This model directly incorporates
the principle of local activation and long-range inhibition which is fundamental to pattern formation in Turing systems, but which is only
indirectly implemented via reaction-diffusion interactions in such systems. We have demonstrated that competitive coupling of spatially
distinct maps results in complex stable patterns even when the site dynamics is trivial, in the sense that the local map has a stable fixed point. 
Further, we have shown that the patterns formed by this mechanism are qualitatively independent of the detailed form of the local
site dynamics and of the spatial geometry. Thus, competitively coupled maps provide a simple and robust model of spatial pattern formation.

B.S. is grateful to Rainer Scharf and Permanand Indic for helpful comments.


\begin{thebibliography}{}
\bibitem{Giereretal} A. Gierer and H. Meinhardt, Kybernetik {\bf 12}, 30 (1972); S. Douady and Y. Couder, Phys. Rev. Lett.  {\bf 68}, 2098 (1992);
A. Koch and H. Meinhardt, Rev. Mod. Phys. {\bf 66}, 1481 (1994); E. Boissonade, E. Dulos and P. De Kepper, in {\it Chemical Waves and Patterns},
eds. R. Kapral and K. Showalter, 221 (Kluwer, Dordrecht, 1995).
\bibitem{Turing} A. Turing, Phil. Trans. R. Soc. {\bf B237}, 37 (1952).
\bibitem{BZ} R. Kapral and K. Showalter (eds.), {\it Chemical Waves and Patterns} (Kluwer Academic Publishers, Dordrecht, 1995); S. Jakubith, H. H. Rotermund, W. Engel, A. von Oertzen and G. Ertl, \prl {\bf 65}, 3013 (1990); G. Ertt, Science {\bf 254}, 1750 (1991); J. Lechleiter, S. Girard, E. Peralta and D. Chapman, Science 
{\bf 252}, 123 (1991); K. Agladze, J. P. Keener, S. C. Muller and A. Panfilov, Science {\bf 264}, 1746 (1994); F. Mertens and R. Imbihl, Nature {\bf 370}, 124 (1994).
\bibitem{Meinh} H. Meinhardt, {\it Models of Biological Pattern Formation} (Academic Press, London, 1982); J. D. Murray, {\it Mathematical Biology}
(Springer, Berlin, 1990); A. Koch and H. Meinhardt, Rev. Mod. Phys. {\bf 66}, 1481 (1994); H. Meinhardt, Nature {\bf 376}, 722 (1995).
\bibitem{Castit} V. Castets, E. Dulos, J. Boissonade and P. De Kepper, Phys. Rev. Lett.  {\bf 64}, 2953 (1990); Q. Ouyang and H. L. Swinney, Nature {\bf 352}, 610 (1991);
K. J. Lee, W. D. McCormick, H. L. Swinney and J. E. Pearson, Nature {\bf 369}, 215 (1994); I. R. Epstein and V. K. Vanag, CHAOS {\bf 15}, 047510 (2005).
\bibitem{Kaneko} K. Kaneko, Prog. Theo. Phys. {\bf 72}, 480 (1984); J. P. Crutchfield, Physica {\bf D10}, 229 (1984); T. Yamada and H. Fujisaka, Prog. Theor. Phys. {\bf 72}, 885 (1984); K. Kaneko, CHAOS {\bf 2}, 279 (1992).
\bibitem{Keel} J. D. Keeler and J. D. Farmer, Physica {\bf 23D}, 413 (1986).
\bibitem{WallKap} I. Waller and R. Kapral, Phys. Rev A {\bf 30}, 2047 (1984).
\bibitem{LJans}  V. A. A. Jansen and A. L. Lloyd, J. Math. Biol. {\bf 41}, 232 (2000).
\bibitem{Jack90} E. A. Jackson, {\it Perspectives in Nonlinear Dynamics, Vol. 2} (Cambridge University Press, 1990).
\bibitem{Roh96} P. Rohani, R. M. May and M. P . Hassell, J. Theor. Biol. {\bf 181}, 97 (1996).
\bibitem{Kap85} R. Kapral, Phys. Rev. A {\bf 31}, 3868 (1985).
\bibitem{Yam83} T. Yamada and H. Fujisaka, Prog. Theor. Phys. {\bf 70}, 1240 (1983).
\bibitem{CMan88} H. Chat\'{e} and P. Manneville, Physica D {\bf 32}, 409 (1988).
\bibitem{CMan92} H. Chat\'{e} and P. Manneville, Europhys. Lett. {\bf 17}, 291 (1992).
\bibitem{Pol93} A. Politi, R. Livi, G. L. Oppo and R. Kapral, Europhys. Lett. {\bf 22}, 571 (1993).
\bibitem{Hara} M. Yokozawa, Y. Kubota and T. Hara, Ecol. Model. {\bf 106}, 1 (1998); M. Yokozawa and T. Hara, Ecol. Model. {\bf 118}, 61 (1999).
\bibitem{DoebKill} M. Doebeli and T. Killingback, Theor. Popul. Biol. {\bf 64}, 397 (2003).
\bibitem{Lateral} H. Meinhardt and A. Gierer, J. Cell. Sci. {\bf 15}, 321 (1974); J. Collier, N. Monk, P. Maini and J. Lewis, J. Theor. Biol. 
{\bf 183}, 429 (1996). 
\bibitem{N1}J. R. S\'{a}nchez, J. Gonz\'{a}lez-Est\'{e}nez, R. L\'{o}pez-Ruiz and M. G. Cosenza, Euro.Phys.J. ST {\bf 143}, 241 (2007).
\bibitem{N2}J. Gonz\'{a}lez-Est\'{e}nez, M. G. Cosenza, R. L\'{o}pez-Ruiz and J. R. S\'{a}nchez, Physica A {\bf 387}, 4637 (2008).
\bibitem{N3}J. Gonz\'{a}lez-Est\'{e}nez, M. G. Cosenza, O. Alvarez-Llamoza and R. L\'{o}pez-Ruiz, Physica A {\bf 388}, 3521 (2009).
\bibitem{N4}J. Hofbauer, V. Hutson and W. Jansen, J. Math Biol. {\bf 25}, 553 (1987).
\bibitem{N5}J. M. Cushing, S. Levarge, N. Chitnis and S. M. Henson, J. Diff. Eqns. Appl., {\bf 10}, 1139 (2004).
\bibitem{May} R. M. May, Nature {\bf 261}, 459 (1976); M. J. Feigenbaum, J. Stat. Phys. {\bf 19}, 25 (1978); M. J. Feigenbaum, J. Stat. Phys. {\bf 21}, 669 (1979).
\bibitem{Davis} P. J. Davis, {\it Circulant Matrices} (Chelsea Publishing, New York, 1994).
\bibitem{Sadoc} J.-F. Sadoc and R, Mosseri, {\it Geometrical Frustration} (Cambridge University Press, Cambridge, 1999); D. Nelson, {\it Defects and Geometry 
in Condensed Matter Physics} (Cambridge University Press, Cambridge, 2002).
\bibitem{SG} D. Chowdhury, {\it Spin Glasses and Other Frustrated Systems} (Princeton University Press, 1987); J. Mydosh, {\it Spin Glasses} (Taylor \& Francis, 1995). 
\end{thebibliography}
\end{document}